\documentclass[10pt,twocolumn,letterpaper]{article}

\usepackage[utf8]{inputenc}
\usepackage[export]{adjustbox}
\usepackage{multirow}
\usepackage{tabularx}
\usepackage{graphicx}
\usepackage{subfigure}
\usepackage{color}
\usepackage{amsmath}
\usepackage{multirow}
\usepackage{rotating}
\usepackage{microtype}
\usepackage{wrapfig}
\usepackage{soul}
\usepackage{cite}
\usepackage{xurl}

\begin{document}

\title{Exploration of Enterprise Server Data to Assess Ease of Modeling System Behavior}
\author{Enes Altinisik, Husrev Taha Sencar\footnote{For further information and questions, please contact Enes Altinisik ($ealtinisik@hbku.edu.qa$) or Husrev Taha Sencar ($hsencar@hbku.edu.qa$).}, Mohamed Nabeel, Issa Khalil, and Ting Yu\\
Qatar Computing Research Institute}
\date{}

\maketitle

\begin{abstract}
Enterprise networks are one of the major targets for cyber attacks due to the vast amount of sensitive and valuable data they contain.
A common approach to detecting attacks in the enterprise environment relies on modeling the behavior of users and systems to identify unexpected deviations. 
The feasibility of this approach crucially depends on how well attack-related events can be isolated from benign and mundane system activities.
Despite the significant focus on end-user systems, the background behavior of servers running critical services for the enterprise is less studied. 
To guide the design of detection methods tailored for servers, in this work, we examine system event records from 46 servers in a large enterprise obtained over a duration of ten weeks.
We analyze the rareness characteristics and the similarity of the provenance relations in the event log data.
Our findings show that server activity, in general, is highly variant over time and dissimilar across different types of servers. 
However, careful consideration of profiling window of historical events and service level grouping of servers improve rareness measurements by 24.5\%. Further, utilizing better contextual representations, the similarity in provenance relationships could be improved.
An important implication of our findings is that detection techniques developed considering experimental setups with 
non-representative characteristics may perform poorly in practice. 

\end{abstract}

\section{Introduction}
Detection of and response to cyber attacks in the enterprise continue to be challenging despite the vast array of security solutions available today.
The scale of the enterprise environment in terms of the number and diversity of end-user devices and the variety of applications and services, coupled with human factors that impact security, creates a very wide attack surface.  
Significant research and development effort has been devoted to build capabilities that help detect, assess, investigate and respond to sophisticated attacks such as Advanced Persistent Threats (APTs).

An important focus of enterprise security has been on developing techniques that combine rule- and anomaly-based 
approaches to detect targeted attacks \cite{dai2013patrol,streamspot,berrada2019aggregating,conan,sleuth,holmes,han2018provenance,unicorn,deeplog,liu2018anomaly,hu2017anomalous,yuan2020time,sharma2020user,corney2011detection,tuor2017deep,rashid2016new}.
These approaches can be divided into two general groups. 
The first group is centered around the actions of a user by developing models representing normal user behavior of interacting with systems \cite{maloof2007elicit,log2vec,sharma2020user}.
These models are essentially driven by baseline parameters obtained from observed user activities on these systems.
Inevitably, the success of such user behavior models depends on three main factors: how well they can capture dynamics of user actions, their ability to correctly identify the context within which the actions are taken, and their capacity to distinguish user behavior from the innocuous background behavior of the system.

The other group of approaches is more system-centric and mainly seeks to developing capabilities that can either identify suspicious patterns in system execution in relation to previously observed attacker actions \cite{holmes,MITREATT,pei2016hercule,milajerdi2019poirot} or 
characterize the normal system behavior through developing models \cite{priotracker,nodoze,unicorn,zengwatson}.
More recently, as part of DARPA's Transparent Computing (TC) program \cite{darpa,darpaGithub} a new wave of techniques based on data provenance analysis has appeared.
Such approaches showed that a causality analysis framework can also be used to detect anomalies in system execution caused by an APT attack.
All the above approaches to enterprise security mainly considered end-user systems in their analyses as they provide a convenient entry point into the enterprise network for attackers.

In principle, such approaches can also be applied to organizational servers that provide services to the whole enterprise, although, their dynamic and central nature requires special consideration.  
In this regard, a thorough understanding of the server environment is key for developing solutions that will be effective in practice. 
Moreover, the effectiveness of such solutions on servers is much more critical than end-user systems as the value of compromising a domain controller or a database server is much higher and accessing servers is among the end-goals of an attacker. 
To address this, the TC program generated much needed datasets in the field that simulated the enterprise environment by also including some security-critical services such as WEB, SSH, and Exchange servers in addition to end-host systems. 
However, the benign state of the systems was generated by running a scripted set of activities on each server. 
One drawback of this is that it makes it relatively easy to model benign behavior as it does not represent the actual server behavior in terms of the amount of workload, types of activity, and variety of applications, thereby creating a favorable scenario for detecting traces of an attack. 

In fact, one of the major obstacles in enterprise security research is the lack of realistic and up-to-date datasets to facilitate the design and evaluation of new methods \cite{kenyon2020public}. 
Given the challenges of obtaining actual enterprise data that can drive research, a feasible solution is to create realistic synthetic data.
This essentially requires accurately profiling the execution behavior of servers to create formal models that can be used in the generation of synthetic datasets. 
As an example, the widely used CERT dataset \cite{cert} used for insider threat detection is aimed at simulating human behavior based on observations made by earlier work on the network and host events in a large enterprise \cite{vishwanath2009swing,wright2010generating,salem2011modeling}. 
The ability to create such datasets with a high degree of realism requires extending such measurements to servers as the most important components of an enterprise network.

With these motivations, in this work we study system audit data obtained from 46 servers in a large organization.
The data under investigation spans a ten-week period and includes various aspects of activities performed by both administrator and standard users on those servers.
Our observations show that servers exhibit quite a high variation in their behavior depending on the services they are running.
In this regard, user logins to systems do not exhibit an apparent pattern on most servers which indicating the difficulty of modeling logon behavior.
Our analysis of the server data with respect to the rareness of observed events and the similarity of the created provenance graphs representing system activity \cite{gehani2012spade,li2021threat} over time further supports this finding.
In terms of system activity, it is determined that on average 90\%, 60\%, 30\%, and \%15 of the registry, file, process, and network events, respectively, are encountered for the first time when event occurrences are evaluated with respect to previous weeks.
The graph similarity between weekly generated provenance graphs, assessed using a histogram similarity measure based on subgraph representations of 2-hop node neighborhoods \cite{unicorn,streamspot}, is found to be on average 29\% with the highest individual value being 65\%.
Overall, this work brings to light a less examined side of server activity in an enterprise.

Our paper is organized as follows. 
In the next section (§ \ref{sec:data}), we provide a description of what our data comprises and discuss the categorization of servers.
We then provide statistics on users' logon activity (§ \ref{sec:overview-logon}) and examine how file, process, network, and registry operations vary across these categories over time across servers (§ \ref{sec:overview-data}).  
Results for the analysis of data in terms of rareness of events (§ \ref{sec:rareness}) and self-similarity  of provenance graphs (§ \ref{sec:similarity}) are provided next (§ \ref{sec:analysis}).
The paper is then concluded with a discussion of findings that may guide further research in this field.

\section{Enterprise Server Log Data}
\label{sec:data}

To understand and study the enterprise server environment, we used event logs containing records of activities carried out by a variety of servers in a large healthcare organization.
The log data was collected over a period of 10 weeks in early 2019 from 46 physical systems.
These servers perform 23 different services that are widely used in and essential for the enterprise environment with some services running on multiple servers such as WEB servers, exchange servers, domain controllers, and database servers.
The collected log data are intended to support auditing and cyber security incident investigations and archived in a central log storage system.  

\subsection{Source and Composition of Logs}

All the servers in the organization run Windows operating system (OS), one of the most commonly used server OS worldwide \cite{serverDist}.
Windows provides two powerful system services to collect event logs, called {\em Event Tracing for Windows} (ETW) and {\em System Monitor} (Sysmon).
ETW is a built-in tracing facility to log kernel, application, and system activities for diagnostic purposes.
It can provide extensive information about the runtime state of a system and might be configured to collect event logs from selected sources.

Sysmon is the other system service focusing primarily on the collection of security-related events that are widely used for threat detection and forensics purposes \cite{SysmonWi3}.
It is provided as a system utility tool by Microsoft and must be installed independently. 
Since Sysmon is mainly developed for security monitoring, its logs provide a richer description of the event context than the Windows event logs provided by the ETW.  For example, a process creation event can be logged by both ETW (event ID 4668)  and Sysmon (event ID 1), however, the hash of the process and command line arguments to the parent process are readily available only in the latter.

The logs collected by these services provide detailed information about user logons, process creations, network connections, file operations, and registry operations. Computing systems can generate huge volumes of log data, and this imposes a significant storage and analysis challenge, especially for large-scale enterprises such as the one under examination.
Therefore, to maintain the long-term investigation capability, event logs deemed not to be related to security incidents are typically filtered out before being sent to the central security information and event management (SIEM) system. 

Our event log data are generated using both ETW and Sysmon and include activities of 18 system administrators and 1,580 standard users performed on these servers during the 10-week period. 
To avoid redundancy in collected event data, both collection services are configured to selectively monitor different kinds of activities. 
In our environment, user logon/logoff events are collected by ETW and all network and registry events are logged via Sysmon. 
Process and file events are monitored using both services. 

The organization also implements filtering of event logs to eliminate mundane or benign system activities and to reduce the amount of data that needs to be stored. 
For this purpose, ETW is configured to collect process activity related to administrators and operations on files located in selected system and application directories.
By contrast, logon and logoff events are collected without any filtering. 
Sysmon events are collected after application of a rule set \cite{SwiftOnSecurity} that is extensively deployed by organizations \cite{HTAT,AHolisti16}.
These rules are mainly designed to improve the visibility of unusual system activities and are derived from incident reports. 
Accordingly, most of the process events commonly related to Windows system as well as those related to regular activities are eliminated, and only non-standard process events are logged.
That is, important process and file activities are largely preserved.  
By contrast, network and registry activities are logged more conservatively as they contribute significantly to the overall volume.
In this regard, network activity is logged only if the path or name of the source process is defined as possibly suspicious or if the destination port is associated with a potentially harmful application. 
Similarly, due to the extensive amount of registry events in normal operation, only write activities on selected registry entries are monitored.

The trade-off imposed by the need to perform long-term analysis and the filtering of event logs ultimately brings with it the risk of missing attack activities and 
limits the forensic utility of data.
However, the need to reduce log data is inevitable in the enterprise environment with possibly thousands of host systems and tens of servers to ensure that records can support future investigations. 
Overall, our enterprise log data includes all the critical process and file-related event logs essential for analysis. 
Further, since the servers are deployed within the enterprise network to provide internal users with well-defined tasks, they inherently have limited external network interactions compared to end-user systems.

\subsection{Categorization of Servers}

An analysis of logs at the server level will not be feasible due to the number of servers. 
Therefore, we consider grouping them to explore their behavior categorically. 
An end-user system's behavior is mainly determined by the diversity of user actions, and a degree of similarity among hosts is expected depending on  
the user's role and the functional unit in the enterprise they belong to.  
In contrast, a server's behavior is differentiated by the service it provides and the interactions with users of this service. 
Since at the service level each server will be more or less unique, the basis of categorization must be the latter. 
From the user perspective, some servers may require application-level authentication whereas others may involve operating-system-level authentication. 
These two modes of operation will have a different reflection on the event logs, with the latter leaving more visible traces on the system execution. 

Hence, considering activities of administrator and standard users, we group servers into three categories as displayed in Table \ref{Servers}.
To differentiate between administrator and standard users, we used process creation events logged by both Sysmon (event ID 1) and 
ETW (event ID 4688) together. In our environment, the latter was configured to only log activity by administrator users, thereby making it possible to identify user roles.
The first group includes 26 servers that mainly run core infrastructure services as well as those that perform central services.
These servers typically require application-level authentication for all users. 
Therefore, standard users do not need to logon to these systems, and user activity can only be discriminated through application logs.
The second category includes task-oriented servers used by a small subset of standard users who connect these systems remotely and run interactive sessions.
This category includes 12 servers that are used for development and testing of different enterprise applications and those that support them.
User activities on these servers are expected to be similar to those on end-user systems. 
The last category includes the remaining eight servers that run mission critical services for the enterprise network.
All users need to frequently interact with these systems, typically, through non-interactive logons.
In common, all these servers are maintained by a group of administrator users that access these servers to perform administration and configuration tasks. 
 
The three categories of servers listed in Table \ref{Servers} (first column) largely overlap with an energy consumption-based categorization of servers with first category servers consuming the least energy followed in order by the servers in the two other categories \cite{uddin2010server}.
It must be added that the 46 servers comprise 23 different services with 1-4 instances for each type of server. 
Those servers with multiple instances in some cases support different applications, such as the  WEB and SQL servers, and in others they are located at different physical locations of the enterprise, such as the domain controllers (DCs) and some development servers.
In this regard, WEB and SQL servers deliver services for different applications and DC and some development servers are deployed at different physical locations of the enterprise.
Finally, it must also be noted that some types of servers appear in different categories as they exhibit different user-access patterns.

\begin{table}[!ht]
	\centering
	\caption{Distribution of servers across three different categories based on access pattern of administrator users and standard users.}
	\label{Servers}
	\resizebox{\linewidth}{!}{
	\begin{tabular}{|c@{\hspace{0.2\tabcolsep}}c|c|c|c|}
	\hline
	\multicolumn{2}{|c|}{Access} & Name & Abbreviation & \# of  \\
	\multicolumn{2}{|c|}{Pattern} & & & Servers\\ \hline
	\parbox[t]{2mm}{\multirow{13}{*}{\rotatebox[origin=c]{90}{Only Admins /}}} & \parbox[t]{2mm}{\multirow{13}{*}{\rotatebox[origin=c]{90}{Category 1}}}
	    &  Systems Group Infrastructure & STG & 2 \\ \cline{3-5} 
	    &&  Structured Query Language  & SQL & 4 \\ \cline{3-5}
	    &&  Key Management Service   & KMS & 1 \\ \cline{3-5}
	    &&  Quality Management Test   & QMTEST & 1 \\ \cline{3-5}
	    &&  Endpoint Security Solution   & ESS & 4 \\ \cline{3-5} 
	    &&  WEB   & WEB & 4 \\ \cline{3-5}
	    &&   SQL Developer (DEV)  & SQLDEV & 2 \\ \cline{3-5}
	    &&  Application   & APP & 3 \\ \cline{3-5}
	    &&  Dynamic Host Configuration Protocol   & DHCP & 1 \\ \cline{3-5}
	    &&  Production  & PROD & 1 \\ \cline{3-5}
	    &&  Backup   & BKP & 1 \\ \cline{3-5}
	    &&  Office Online Server & OOS & 1 \\ \cline{3-5}
	    && SharePoint (SP) Production Application & SPPRODAPP & 1 \\ \hline
	\parbox[t]{2mm}{\multirow{9}{*}{\rotatebox[origin=c]{90}{Admins \& Some Users /}}} & \parbox[t]{2mm}{\multirow{9}{*}{\rotatebox[origin=c]{90}{Category 2}}}
	    &   CISCO Performance Visibility Manager  & CPVM & 1 \\ \cline{3-5}
	    &&  SharePoint Developer    & SPDEV & 3 \\ \cline{3-5}
	    &&  User Acceptance Testing  & UAT & 1 \\ \cline{3-5}
	    &&   WEB Data Management   & WEBDM & 1 \\ \cline{3-5}
	    &&  Backup   & BKP & 1 \\ \cline{3-5}
	    &&  Production Web   & PRODWEB & 2 \\ \cline{3-5}
	    &&  Local Administrator Password Server   & LAPSVR & 1 \\ \cline{3-5}
	    &&  Production  & PROD & 1 \\ \cline{3-5}
	    &&   SharePoint Production Application  & SPPRODAPP & 1 \\ \hline
	\parbox[t]{2mm}{\multirow{4}{*}{\rotatebox[origin=c]{90}{All Users /}}} & \parbox[t]{2mm}{\multirow{4}{*}{\rotatebox[origin=c]{90}{Category 3}}}
	    &  Mail Exchange   & EX & 2 \\ \cline{3-5}
	    &&  System Center Configuration Manager & SCCM & 1 \\ \cline{3-5}
	    &&  Domain Controller & DC & 4 \\ \cline{3-5}
	    &&  File Server & FILESVR & 1 \\ \hline
	\end{tabular}
	}
\end{table}

\section{Overview of Server Activity}

We provide an overview of user activity on servers in terms of the number and duration of logon sessions as well as how the number of process creations, file operations, network connections, and registry modifications varies across servers over time.

\subsection{Logon Statistics}
\label{sec:overview-logon}

Since a server must respond to user actions and requests, the access behavior of users determines the level of system activity.  
Users can log on to servers through many methods depending on what task they will perform.
However, unlike in end-user systems where most logons are performed locally, servers are mostly accessed from the network.
Our examination of the logon types revealed that almost all server logons are network logons.

The logon behavior of users of end-user systems is successfully utilized to detect attacks that leverage compromised credentials or insider threats  \cite{log2vec, bohara2017unsupervised,siadati2017detecting,yuan2020time}.
These approaches need to be further complemented with user logon patterns to servers. 
Hence, to evaluate the logon behavior, we compute three measures from the server logon records of users.
In doing this, we excluded all logons that were made by service accounts (such as a Web server account that regularly logs on to an SQL server) and only included logons of actual users in the enterprise.
These include the number of times a specific user logs on to a server on average per week, the average number of distinct users who logon to each server per week, and durations between logon and logoff.
Since system activity may vary significantly from day to day, we performed the analysis at the weekly resolution where activity is expected to remain relatively stable across weeks.

Values obtained for each measure by averaging across all servers in the three categories are given in Table \ref{LoginStatistics2019}.
Most notably, these results show that in the first two categories, administrators log on to servers around once a week with each logon session lasting for 7-17 minutes.
For the third category, this number increases to almost 34 logons per day, though for shorter durations.
This trend also holds true for standard users. 
This may be explained by the more central role of this group of servers in daily activities, which includes access to file and exchange servers. 
It must also be noted that the number of logons to servers is typically much higher than end-user systems. 
This can be attributed to the fact that these are mostly non-interactive logons performed by user applications on behalf of the user to access shared resources and use common services.
Another interesting finding is that administrators log on to these servers more frequently than standard users, indicating that this category of servers requires more regular monitoring and maintenance.

\begin{table}[!]
	\centering
	\caption{Average Weekly Logon Statistics for Three Server Categories}
	\label{LoginStatistics2019}
	\resizebox{\linewidth}{!}{
    	\begin{tabular}{|c|c|c|c|c|c|c|}
    	\hline
    	\multirow{2}{*}{\shortstack{Server \\ Type}} & \multirow{2}{*}{\shortstack{User \\ Type}} & Avg. No. Logons & Avg. No. Total & Avg. Logon   \\
    	 &  & Per-User & Distinct Users  & Duration (mins)  \\ \hline
    	\hline
    	Category  & Standard & - & - & -\\ \cline{2-5}
    	1  & Admin & 9.68 & 0.84 & 7.82\\ \hline
    	\hline
    	Category  & Standard & 37.7 & 1.85 & 17.57\\ \cline{2-5}
    	2 & Admin & 11.95 & 0.99 & 21.38\\ \hline
    	\hline
    	Category & Standard & 173.25 & 1,219 & 4.07 \\ \cline{2-5}
    	3 & Admin & 247.15 & 14.99 & 4.02\\ \hline
    	\end{tabular}
	}
\end{table}

\begin{figure}[!]
\centering
    \includegraphics[width=1\columnwidth]{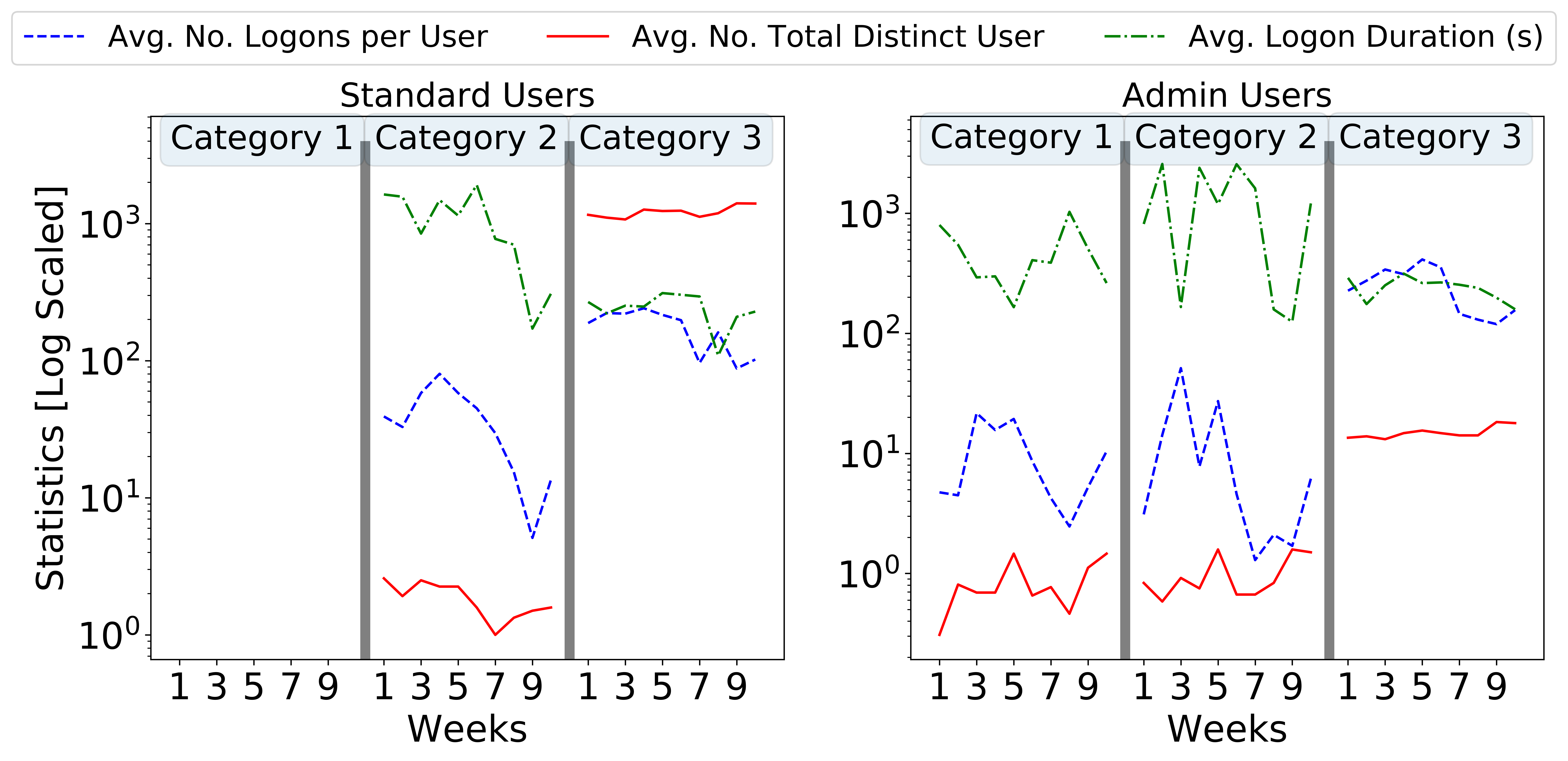}
  \caption{Weekly logon statistics over the weeks.}
  \label{fig:Parsed2019}
  \vspace{-0.2in}
\end{figure}

Fig. \ref{fig:Parsed2019} displays the variation of these measurements over time.
The highest variance is observed in the second category since these servers are mainly used for innovation activities.
This is followed by the first category of servers that only includes administrator logons, and measurements show that their activities on these servers show significant variation over time. In contrast, the variance is relatively low for the third category as a very large number of users regularly access these servers. 
These variations concerning administrators and standard users in the first two categories can be attributed to the varying nature of their daily tasks.
This essentially means user behavior cannot be solely modeled based on historic data in a reliable manner. 
Therefore, attack detection systems need to incorporate other auxiliary information about user activities when assessing abnormal behavior.
We also explored how logon statistics vary during work and off-work hours. 
We determined that user logons don't show a significant variation during the day for servers in the last category.
For the other two categories, most of the logon events take place during work hours as expected.

\subsection{System Activity}
\label{sec:overview-data}

\begin{figure}[!]
  \includegraphics[width=0.9\columnwidth]{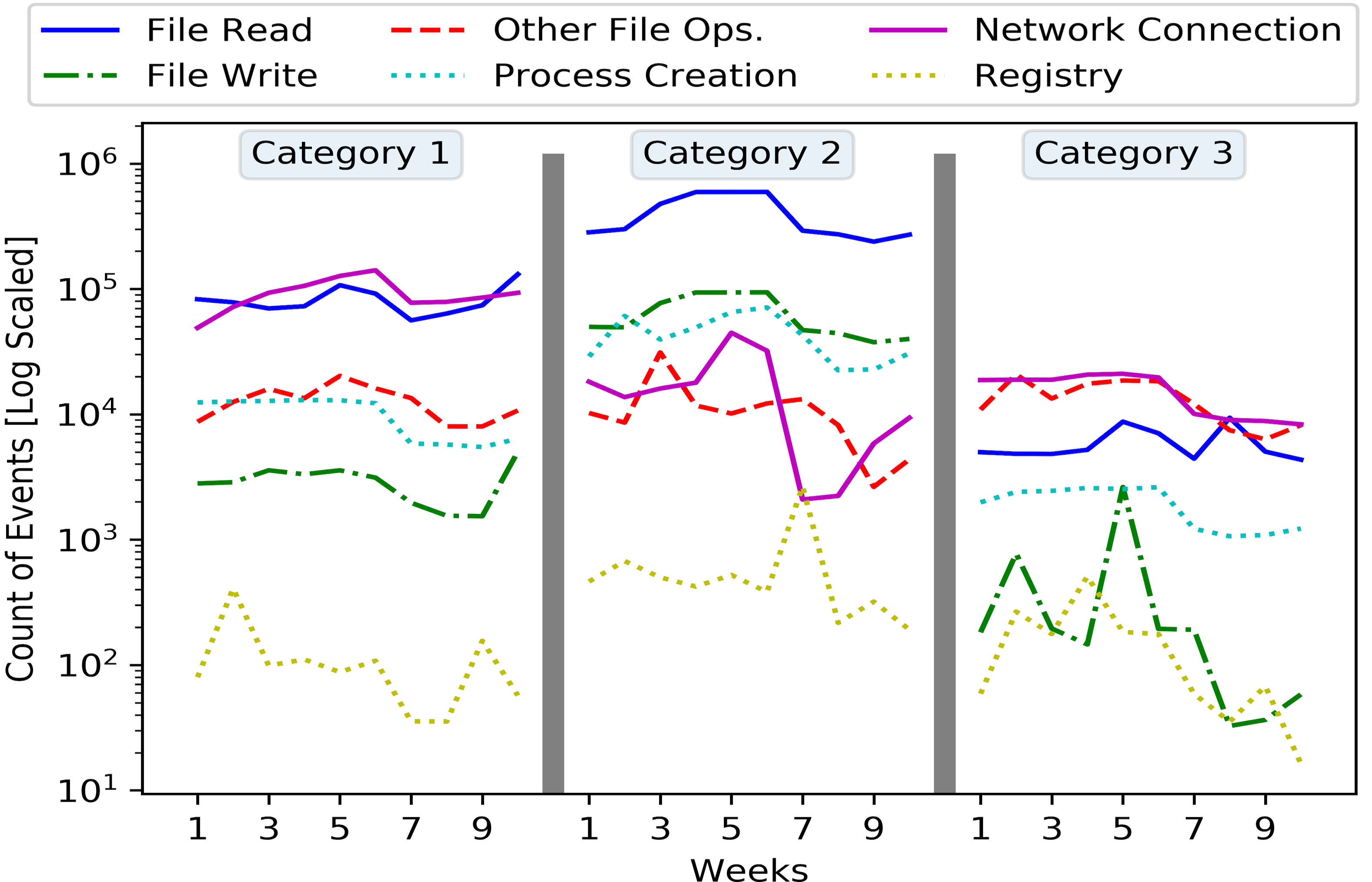}
  \caption{Sever activity measured in terms of process, file, network, and registry operations across weeks}
  \label{fig:othersAcrossTime2019}
  \vspace{-0.2in}
\end{figure}

Next, we examine the number of operations related to processes, files, network connections, and registries performed on each server.
All measurements concerning system activity are obtained considering all users in the system, including system accounts as well, to have a more holistic view.
Fig. \ref{fig:othersAcrossTime2019} shows the variation in system activity across the three categories of servers.
In all cases, file operations dominate the overall activity.
This is followed by network operations in the servers comprising the first and the third categories.
In the second category of servers, however, file operations are followed by process creations in line with strong developer activity performed on these servers.
We also observe that in the third group, file operations and network connections closely follow each other. 
In addition, the least number of process creation events are observed in this category of servers. 
This indicates the service oriented nature of these servers with few out-of-routine processes running on them.
Finally, we determine that except for the second category of servers, system activity does not show significant variation across day and night. 
For the third category of servers, this indifference is most likely due to the inclusion of activity related to system accounts which perform most of the tasks on the system.

\section{Analysis of Audit Data}
\label{sec:analysis}
Enterprise system data have been analyzed for two main aspects. 
The first is a user-centric analysis to mainly detect insider threats and system compromises. 
This approach uses features obtained from user activities such as logons, file accesses, and network connections to create an individual or role-based baseline model and to capture large deviations \cite{wang2012detection,gavai2015detecting,rashid2016new,tuor2017deep} and outlier events \cite{eldardiry2014multi,le2018evaluating,ahmed2019role2vec,log2vec}. 

The other aspect is system-centric and involves the analysis of system logs to identify anomalous events at the system level. 
One approach in this direction is based on learning the complex causal relationships between log entries through dependency-analysis methods \cite{deeplog,shen2018tiresias,du2019lifelong}.
To facilitate better causality assessment, more recently data provenance analysis gained significant attraction.   
In this approach, system logs are parsed into provenance graphs to analyze the relationship between system events and to create a more holistic view of system execution \cite{king2003backtracking}.
Since a graph representation allows a better basis for contextual analysis, several methods have been proposed to detect anomalous events in provenance graphs either through identifying rare occurrences \cite{priotracker,nodoze,ProvDetector} or based on clustering of sub-graph sketches for outlier detection \cite{streamspot,unicorn}.

Although most of these analysis methods are mainly developed considering end-user systems, they can in principle be applied to server data.  The server environment, however, provides limited visibility of overall user activities as compared to end-user systems, and thus, server data is of lesser utility for 
detection of anomalies in user behavior when considered alone. 
Therefore, in our exploration, we adopt ideas from provenance and causality analysis approaches to evaluate the variation in server activity over time and across servers.

We must note here that although our enterprise deploys several state-of-the-art host- and network-based security solutions to protect against attacks, the collected server logs may nevertheless include attack events.
To further validate this, we applied sigma signatures \cite{sigma} to event logs from all servers to verify that no attack events were present.
Although it is possible that these measures missed some attack-related events, it is very unlikely that all the servers would be affected by such activity during the whole duration of data collection.
In this regard, our analysis should be expected to provide the general degree of variability and self-similarity of activity in servers.  
The inherent consistency in activity is particularly relevant in the context of detecting changes in system behavior as any change below the level of normal variation cannot be reliably detected. 
This is an important consideration in detecting APT type attacks which are often slow and stealthy and usually evade existing security solutions.

\subsection{Representation of Log Data}
Our analysis starts by mapping server log data into a provenance graph representation.
System audit log events, whether obtained through Sysmon or ETW, follows a basic form that can be simplified as a 3-tuple ({\em subject, relation, object}) where the {\em subject} is a process, {\em object} may be another process or one of the files, registries, or networks sockets, and the {\em relation} is the system call that shows the interaction of subject with objects, such as read, write, or modify, depending on the object's context.
A provenance graph essentially represents the execution of a system by mapping such 3-tuples representing events into a heterogeneous graph where nodes represent objects or subjects and edges are the relations \cite{king2003backtracking,streamspot,barre2019mining}. 
These nodes are further annotated with attributes available in the event logs, such as timestamps, process IDs and port numbers. 
On the graphs, processes and files are represented by their paths, network objects by their source and destination IPs, and registry objects by their keys and subkeys.
Also, to eliminate server specific information, host and user names are removed from paths of processes and files by replacing them with generic identifiers.
The edges in the graph are directed to show the direction of information flow, and the graph is acyclic because edges are added only when new nodes are created. 

\subsection{Rareness of Events}
\label{sec:rareness}

Rare events in system execution are typically indicative of a problem in the system. 
Therefore, the rareness of system events has been used to detect potential anomalies in system behavior by prioritizing analysis of those events.
The feasibility of this approach crucially depends on the assumption that the number of events that occur infrequently is few.
That is, encountering many rare events in a system curtails the effectiveness of this approach in practice. 
With this perspective, we use the occurrence frequency of events as one measure to profile server activity.  

Rareness of an event can be defined in several ways. 
All definitions, however, rely on a historic time period that reflects the normal operation of a system to compute a baseline for the commonly occurring events. 
\cite{tariq2011identifying,priotracker,nodoze,ProvDetector}
For more reliable assessment of rareness, the event context can also be incorporated by assessing the fan-in and fan-out activity related to each {\em subject} and {\em object} of an event on a provenance graph \cite{priotracker,nodoze, ProvDetector}. 
To exploit the presence of multiple similar hosts and to better define normality, the evaluation can be performed also taking into account the number of hosts that encountered a particular event \cite{priotracker, ProvDetector}.

In our examination of the data, we adopted these approaches of these works.
However, since the log data is partly filtered, the complete provenance graphs of systems cannot be obtained. 
This may prevent correct evaluation of the context of an event as some nodes and edges in the graph are missing.
Therefore, so as to not introduce a bias in rareness measurements, event frequencies are not weighted using fan-in and fan-out structures of nodes in the graph.
It must, however, be noted that filtering does not affect the frequency of observed events.

\def\cpar{\hss\egroup\line\bgroup\hss}
\begin{figure*}[!] 
  \begin{minipage}[b]{0.33\textwidth}
    \centering
    \includegraphics[width=0.95\columnwidth,left]{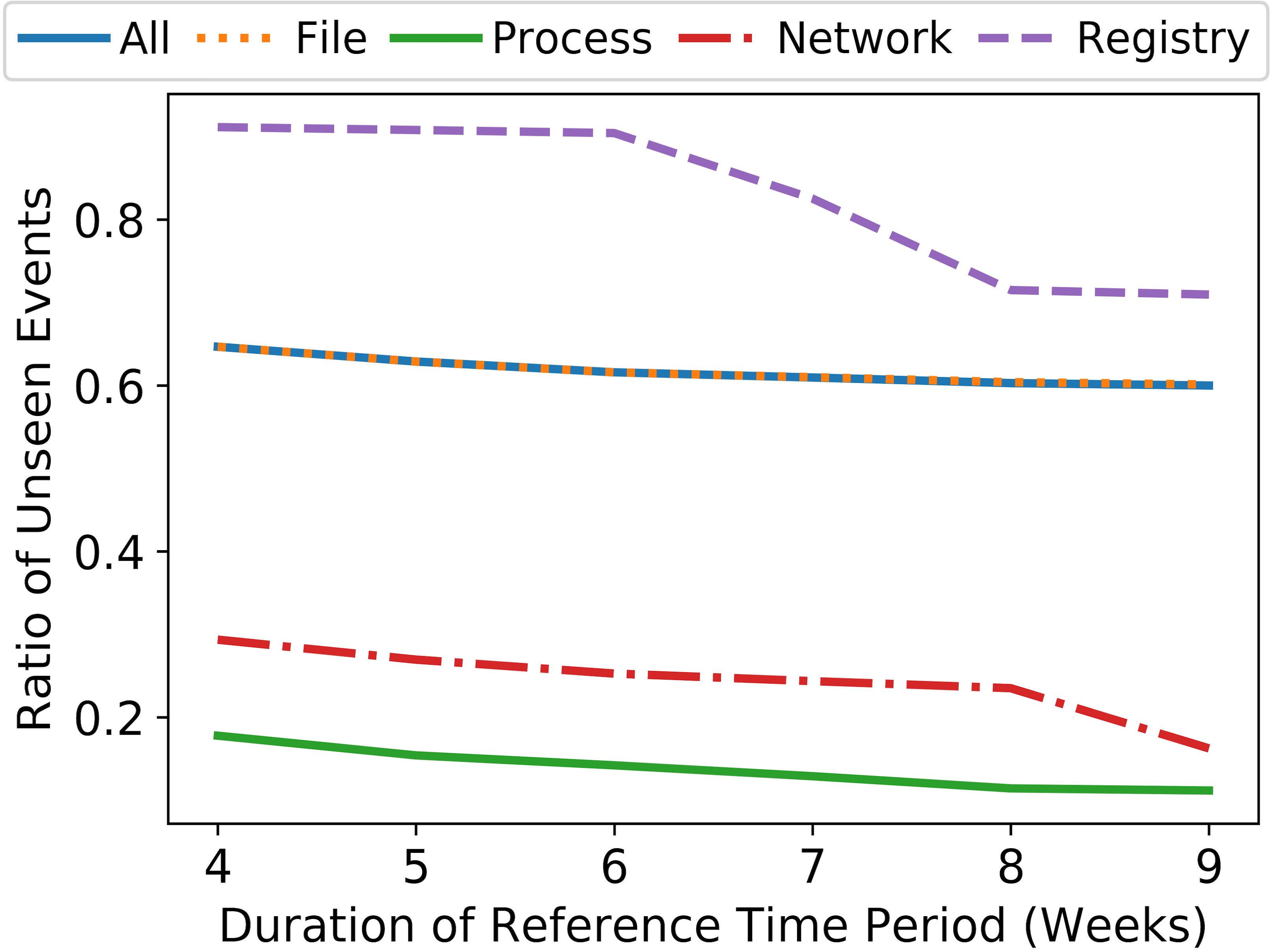}
    \centerline{(a)}
    \label{fig:PrioTrackerAVG}
  \end{minipage}
  \begin{minipage}[b]{0.33\textwidth}
    \centering
    \includegraphics[width=0.95\columnwidth]{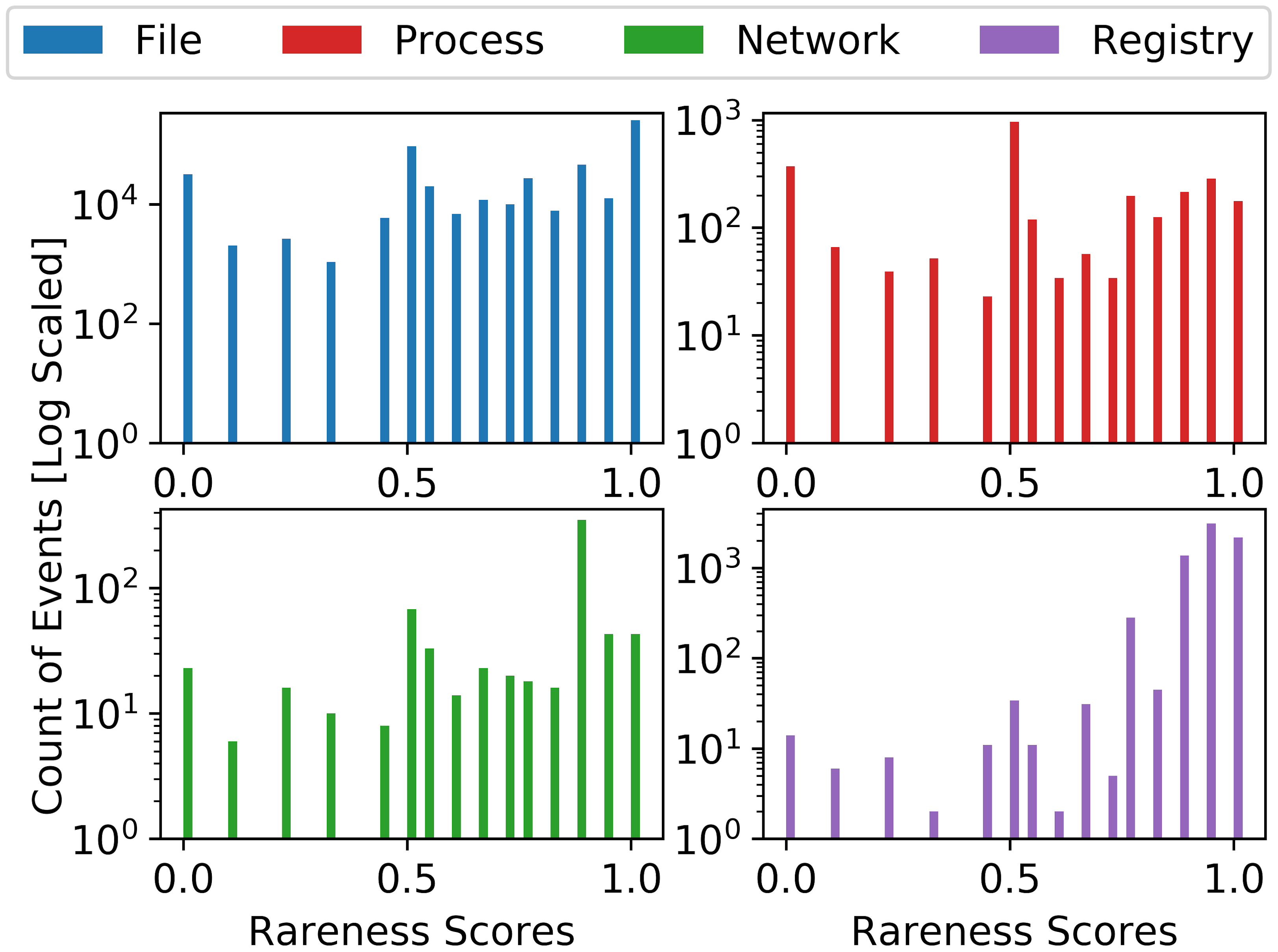}
    \centerline{(b)}
    \label{fig:PrioTrackerDist}
  \end{minipage}
    \begin{minipage}[b]{0.33\textwidth}
    \centering
    \includegraphics[width=0.95\columnwidth,right]{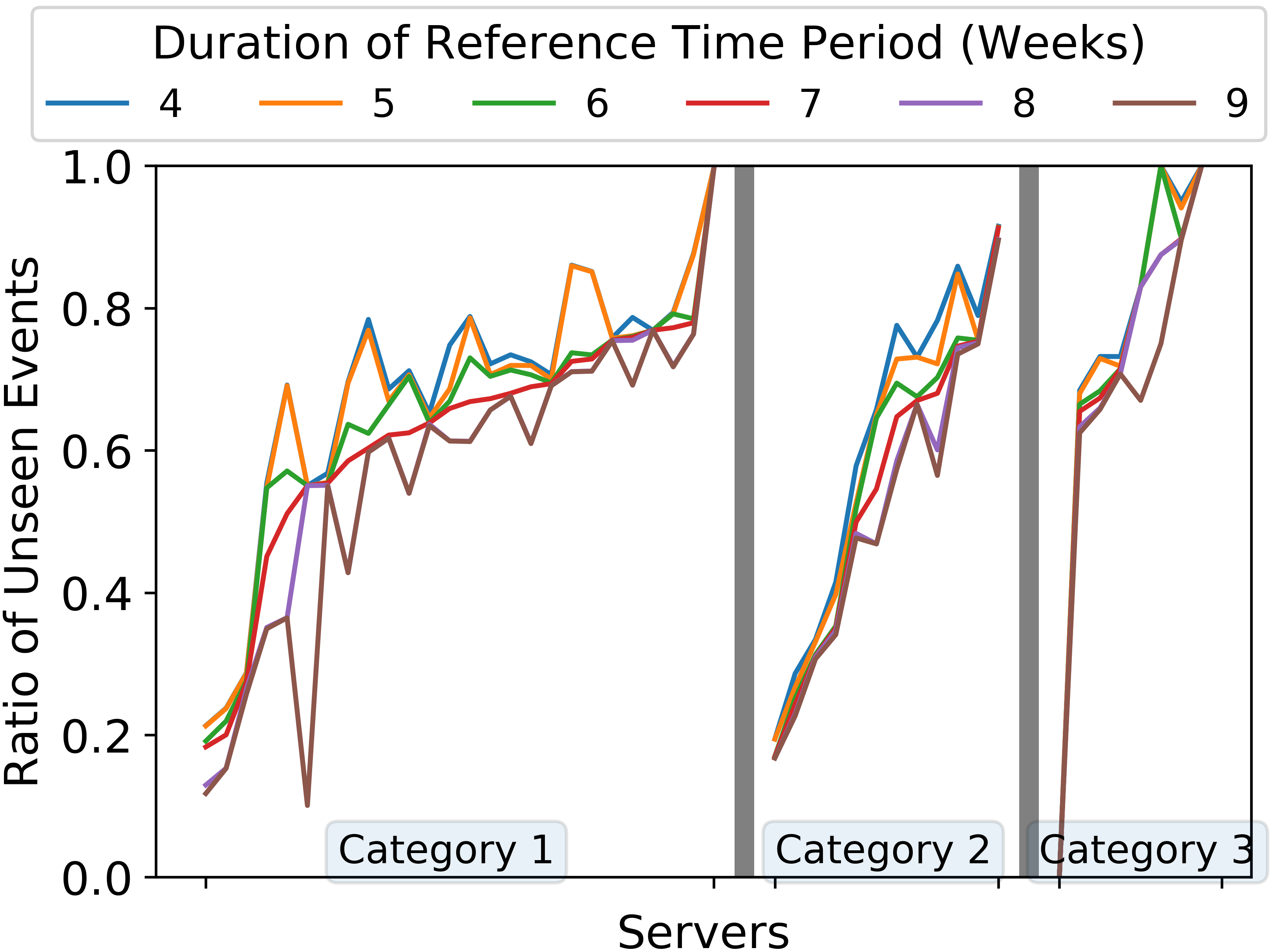}
    \centerline{(c)}
    \label{fig:PrioTrackerDistGroup}
  \end{minipage}
  \vspace{-0.2in}
\caption{Rareness scores computed based on event occurrences observed during the reference time period. 
(a) The ratio of previously unseen events to all events for different types of events when the reference time period varies between 4 and 9 weeks. The scores reflect averages obtained by testing over all weeks.
(b) Distribution of the rareness scores when the first nine weeks are used as a reference window and the tenth week is used for testing. 
(c) The ratio of previously unseen events to all events for each of the 46 servers in the three categories.
}
\label{fig:PrioTracker}
\vspace{-0.2in}
\end{figure*}

\subsubsection{Event Frequencies}
The first measure calculates the rareness score of an event as the inverse of the number of time epochs, such as weeks, in which this event occurs on every server running the same type of service during a designated reference time period \cite{priotracker}.
For this, the data from 46 servers are divided into 31 groups based on the service they run and their physical location in the organization, obtained from server naming conventions of the enterprise. Than the rareness score of an event is computed as
\begin{equation}
    r(e)=1-\frac{\sum_{i=1}^{ W}\sum_{j=1}^{S} {\mathcal I_e}}{\sum_{i=1}^{ W}\sum_{j=1}^{S} {1}}
    \label{eq:rareness1}
\end{equation}
where $W$ in the first sum refers to the total number of weeks in the training period; $S$ in the second sum represents the number of servers in a group taking the value of one or two in our data; and $\mathcal I_e$ is an indicator function that takes the value of $1$ if an observed event occurred at least once during week $i$ on server $j$.  
It must be remembered here that the 46 servers provide 23 different services with up to four servers running the same service. 
In our grouping, servers operating at different physical locations of the enterprise are held separate as they serve different organizational units.
We further discuss in §\ref{sec:discussion} how observed behavior changes depending on the grouping of servers.

Here, regardless of the frequency of an event in a week, it is only counted once since repetitive activities will yield low rareness scores even though they may appear only during a very short duration of time.
For each week from the fifth to the tenth, the rareness scores of events are calculated while the data from all previous weeks are set apart to serve as a reference model.

Fig. \ref{fig:PrioTracker} provides the rareness scores obtained from all servers in a combined manner. 
The ratio of previously unseen events, {\em i.e.,} those with rareness score of $1$, to the overall number of events are given in Fig. \ref{fig:PrioTracker}(a) considering different event types for an increasing reference time period.  
To better characterize this trend, we repeated the measurements on each of the ten weeks while treating the neighboring weeks as the reference time period and obtaining an average.
For a given week, when the number of preceding weeks was less than what is needed for the reference period, data from the upcoming weeks were used.
As can be seen in this figure, over a training period of four to nine weeks ($W$ in Eq. \ref{eq:context}), a great majority of file and registry events have not been seen during the reference time period and around 15\% of the network events and 30\% of the process events remain quite rare. 
Since during its operation, a system is expected to access many different registries and files, this finding is reasonable.

What is more noteworthy is that increasing the reference time period to several weeks causes only a very slow and almost linear reduction in the ratio of rare events. It must be noted, however, that observed rare event ratios would have been much lower if event filtering was not applied.
Elimination of commonly occurring and mundane system tasks causes an overall decrease in the number of events, thereby increasing the visibility of rare occurrences.
Moreover, filtering rules are mainly tuned to capture and log suspicious events, which results in relatively high numbers of rare events.

Therefore, the number of events with high rareness scores, rather than the ratio, is more relevant in practice as it is expected to correlate with the number of cases that will potentially require further investigation. 
Event rarity distributions for different event types obtained from all servers are given in Fig. \ref{fig:PrioTracker}(b).
These rareness scores are computed by retaining event logs of the first nine weeks as reference and assessing if and how frequently events of the last week are observed during this training period. 
Measurements show that events with high rareness scores constitute a large part of all events with those relating to file operations forming the majority.
Accordingly, there are more than 250K file operations that were not encountered in previous weeks.  
A more critical aspect concerns the process activity as it is central to the analysis of system behavior and anomaly detection. 
We determine that previously unseen processes are only around 200 which ultimately is a reasonable number for investigation.

We further examine to see how rare events are distributed across servers.  
Fig. \ref{fig:PrioTracker}(c) shows the ratio of rare events with a score of one to all events for increasing duration of the reference period.
As can be seen in this figure, some servers are more prevalent in their behavior.
In fact, our analysis revealed that 89\% of the unseen events are sourced from five servers that perform many file operations, including a file server, a backup server, and three servers related to deployed endpoint security solution.
In contrast, we determined that task specific servers such as SQL or WEB servers yield very low rareness scores.
Overall, these results show that using file or registry events at server side is going to induce a large variation to the model 
as more than 60\% of their events are new in the following week, whereas this number is quite low for process and network events. 
So, when modeling behavior, changes related to those latter events should be more substantially reflected in the evaluation methodology.

\subsubsection{Contextual Rareness}
\begin{figure}[htbp]
  \begin{minipage}[b]{0.495\columnwidth}
    \centering
    \includegraphics[width=1\columnwidth]{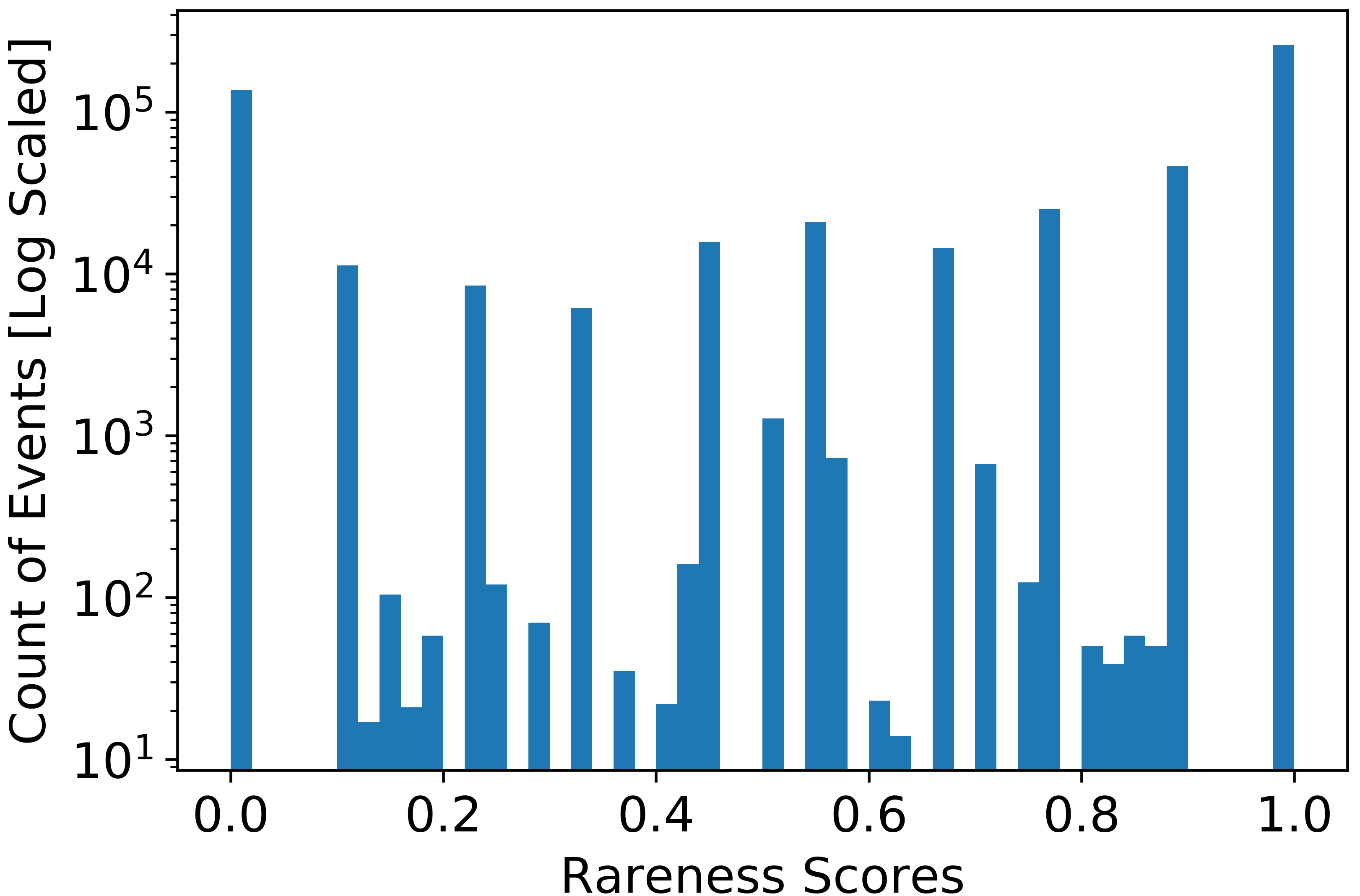}
    \centerline{(a)}
  \end{minipage}
  \begin{minipage}[b]{0.495\columnwidth}
    \centering
    \includegraphics[width=1\columnwidth]{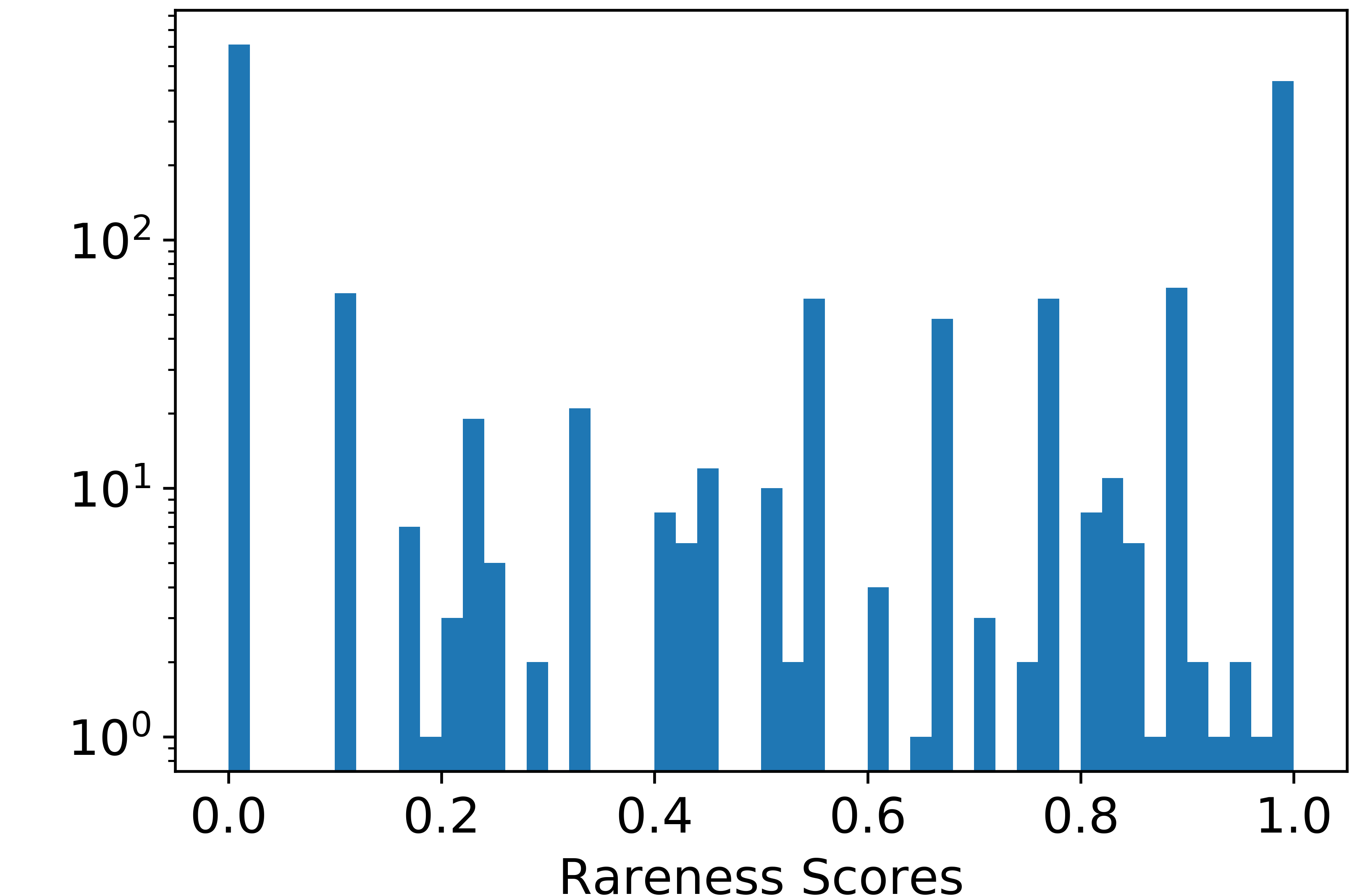}
    \centerline{(b)}
  \end{minipage}
  \caption{Distribution of rareness scores for (a) events and (b) event chains in tenth week when using the first nine weeks as the reference time period.}
  \label{fig:NoDoze}
\end{figure}

As an alternative to measuring the frequency of an event, {\em i.e.,} the 3-tuple of {\em (subject, relation, object)}, among all events that occurred in a server, we also evaluated rareness of an event in the context of similar events with matching {\em(subject, relation)} couple. 
In other words, we assess how commonly a process performs a given task in general as opposed to on a specific object.
This measure computes occurrence probability $P(e)$ of an event $e$ within a more specific context as the ratio of the number of weeks a specific event is seen to have occurred to the number of weeks the particular subject performs the same relation on all objects during the training period \cite{nodoze}.
This occurrence probability, $P(e)$, is subtracted from $1$ to obtain a rareness score $r(e)$ thereby assigning higher scores to increasingly rare events as
\begin{equation}
    r(e)=1- \frac{\sum_{i=1}^{W}\sum_{j=1}^{S} \mathcal I_e}{\sum_{i=1}^{W}\sum_{j=1}^{S} \mathcal I_{e'}}
    \label{eq:context}
\end{equation}
considering a $W$-week training period and assuming $S$ servers are present in the same group.
Similar to Eq. (\ref{eq:rareness1}), the indicator functions $\mathcal I_e$ and  $\mathcal I_{e'}$, respectively, denote the occurrence of a specific event defined as a 3-tuple and any of similar events represented by the corresponding 2-tuple, respectively.
Fig. \ref{fig:NoDoze}(a) shows the distribution of obtained rareness scores for event data collected in the $10^{th}$ week while using the previous nine weeks as the reference time period. 
This measure also shows that there are many events with very high rareness scores.
More specifically, this indicates that many processes are very active and touch many objects in all weeks. 

To add better contextual information, one can alternatively examine event chains with low occurrence probability \cite{nodoze}. 
For this, we examined rareness scores of chains including two connected events that can be obtained in terms of event occurrence probabilities similar to Eq. (\ref{eq:context}), {\em i.e.,} $1-P(e_1)P(e_2)$ for two arbitrary events $e_1$ and $e_2$. 
Fig. \ref{fig:NoDoze}(b) displays the corresponding distribution of rareness scores. 
Accordingly, the number of event chains with rareness score higher than $0.90$ is determined to be 441, out of which 434 were completely unseen. 
This indicates that taking a chain of events into account yields a more favorable context for analysis.

\subsection{Similarity of Provenance Graphs}
\label{sec:similarity}
\begin{figure}[htbp]
 \centering
  \includegraphics[width=0.7\columnwidth]{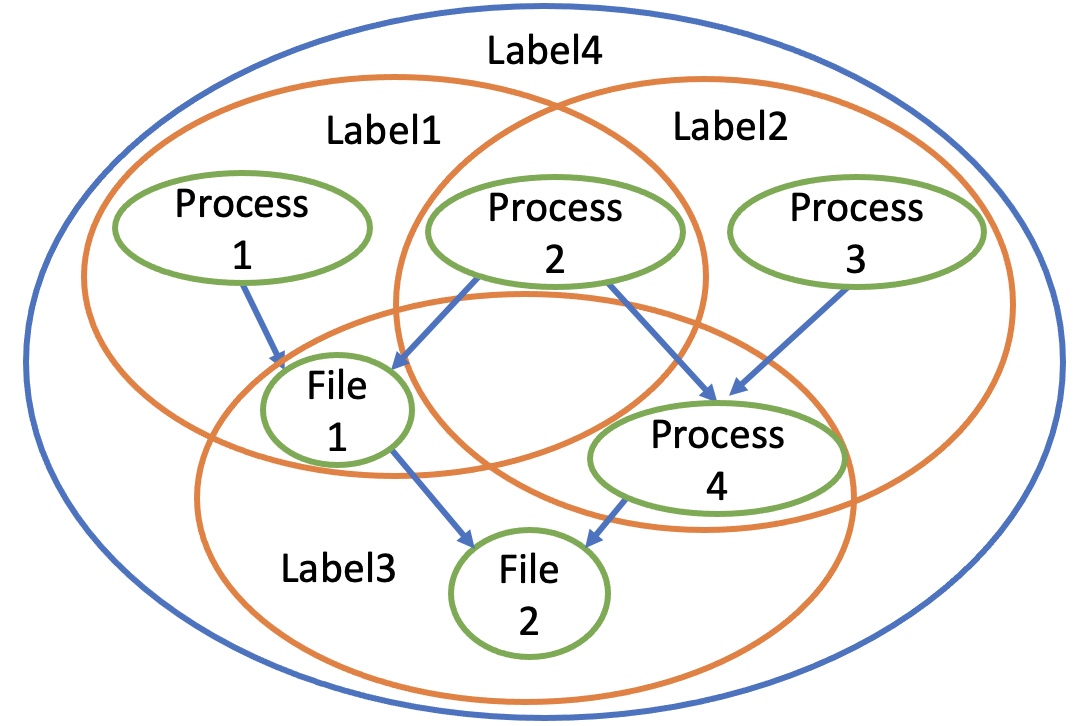}
  \caption{Depiction of how subgraph representations are obtained. The green circles represent nodes or 0-hop node neighborhoods, 
  orange circles represent 1-hop node neighborhoods, and the 2-hop node neighborhood is represented by the blue circle.}
  \label{fig:hop}
\end{figure}

Another approach to analyze server behavior is based on evaluating the structural similarity of the provenance graphs generated from event logs of a server.
The change in provenance graphs over time provides another aspect of the variability of a server's behavior. 
For this, we create a histogram representation of provenance graphs and assess the degree of similarity among those histograms.
In order to capture contextual information of each node in the graph, the subtrees of each node at varying depths are considered \cite{unicorn,streamspot}.
To this end, each node is combined with its incoming edges in a time-sorted manner along with their source nodes into 
a compact representation.
This process is repeated iteratively until reaching the desired path depth that covers larger neighborhoods rooted at a node as demonstrated in Fig. \ref{fig:hop}.
Overall, this yields a subgraph representation that can be assigned labels to create a histogram characterizing a server's behavior. 
To evaluate the similarity among histograms, we use the normalized min-max similarity \cite{cormode2005improved} as a measure.

\begin{figure}[!ht]
  \includegraphics[width=\columnwidth]{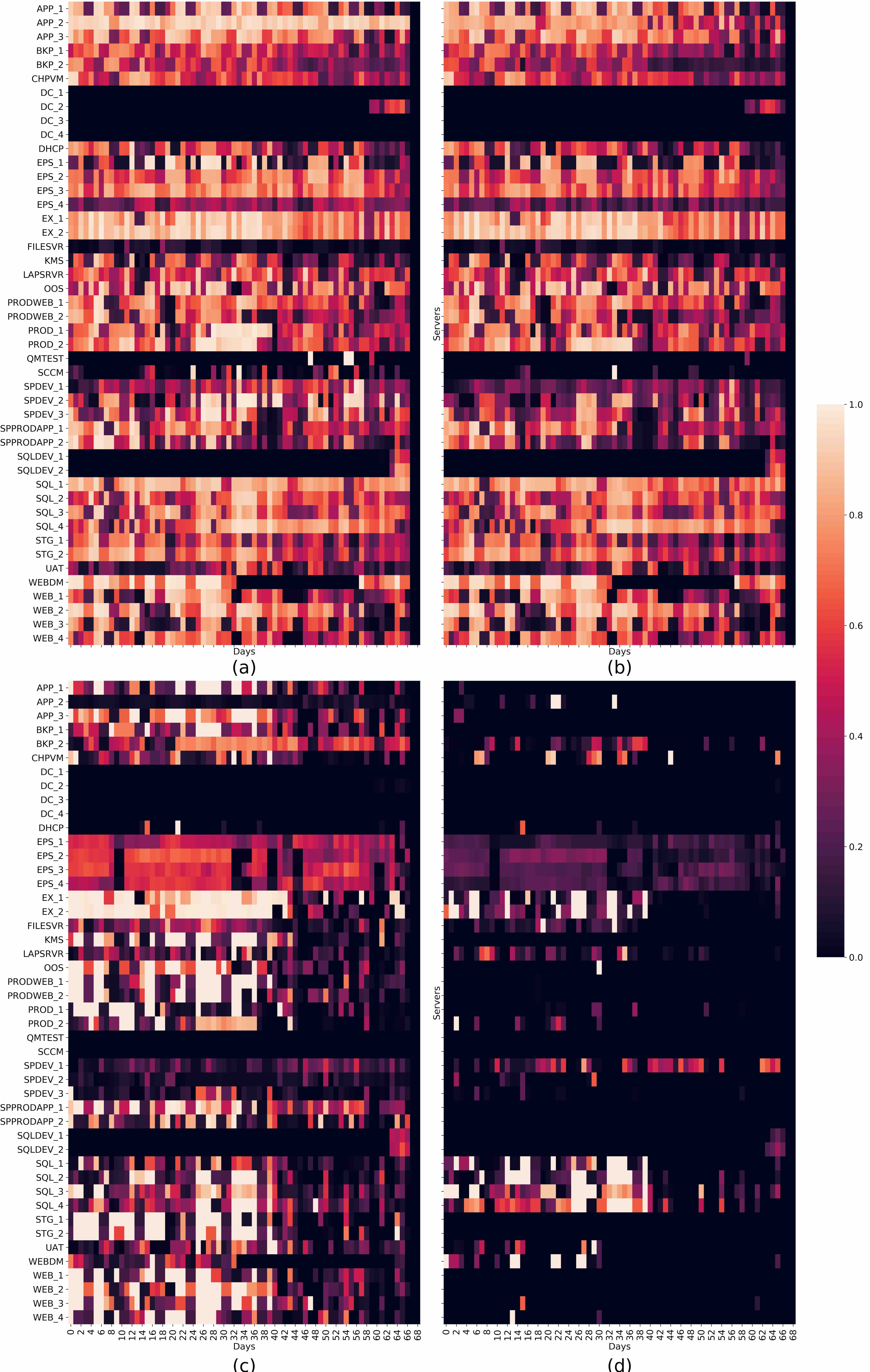}
  \caption{Daily min-max similarity for (a) 0-hop, (b) 1-hop, (c) 2-hop, and (d) 3-hop node neighborhoods of all servers.}
  \label{fig:unicornDaily}
\end{figure}

\begin{figure}[!ht]
  \includegraphics[width=\columnwidth]{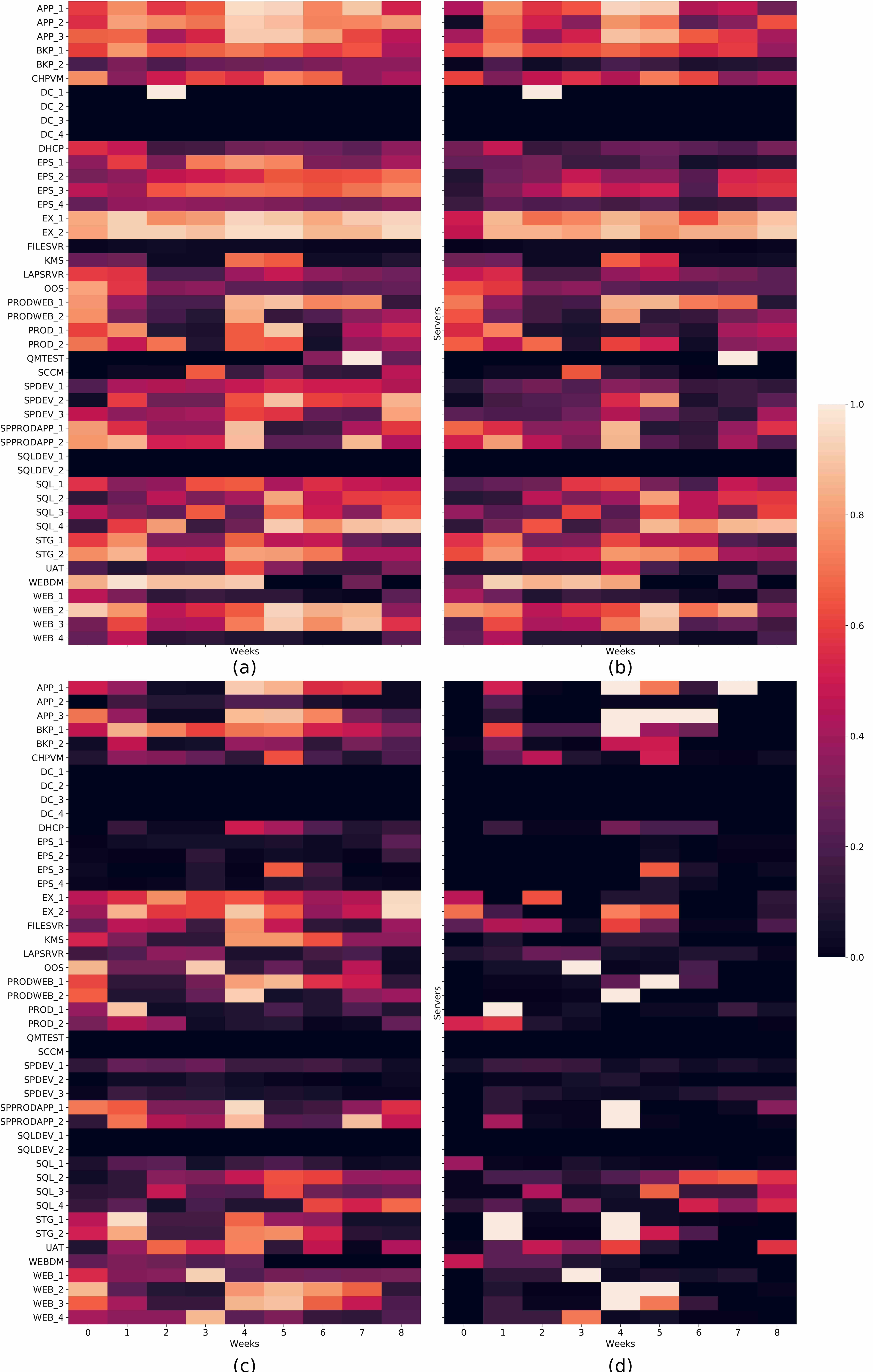}
  \caption{Weekly min-max similarity for (a) 0-hop, (b) 1-hop, (c) 2-hop, and (d) 3-hop node neighborhoods of all servers.}
  \label{fig:unicornWeekly}
\end{figure}

We obtained these subgraph representations spanning  0- to 3-hop node neighborhoods and created histograms of daily and weekly activities of servers.  
Computed similarities between histograms created for subsequent days and weeks are given, respectively, in Figs. \ref{fig:unicornDaily} and \ref{fig:unicornWeekly}.
Accordingly, average similarity considering 0-hop and 1-hop node neighborhoods over all servers shows a similar trend, and for the 1-hop case the similarity at the weekly level is measured to be 
around 40\% which is in line with our earlier findings in Figs \ref{fig:NoDoze}(a) and \ref{fig:PrioTracker}(b) which show that 47\% of events encountered every week have a rareness score of one.

In the case of 1- to 3-hop neighborhoods, however, measured similarities decrease for most of the servers. 
It must be remembered here that due to the filtering of audit data, we can only generate partial system provenance graphs.
Hence, for increasing hop counts the number of labels in the histograms decreases for some of the servers. 
To more reliably observe the similarity of system activity, we identified eight servers \footnote{These servers include four database servers (SQL), a file server (FileSVR), two Exchange servers (EX), and an authentication server (LapSVR).} with a large number of labels computed both at the daily and weekly levels.
Table \ref{MinMAxSim} provides the average similarity scores obtained from multi-hop neighborhoods considering all the servers and the selected eight servers.
It can be seen that as more contextual information is captured (with each histogram label corresponding to a larger neighborhood of nodes in the provenance graph) the similarity unsurprisingly decreases.
Therefore, when more static environments are considered, where the background behavior is more stationary such as the Engagement datasets of the DARPA's TC Program,
larger hop-count neighborhoods may be more preferable as dissimilar events will be more visible. 
For environments like ours, using a low hop-count yields more similarity, thereby making analysis easier.
It is also found that the eight servers exhibit a more similar behavior than others. 
This can be attributed to the fact that these servers perform more regular tasks.

\begin{table}[!ht]
	\centering
	\caption{Average Min-Max Similarities for Subgraphs for Varying Sizes in Hop-Counts}
	\label{MinMAxSim}
	\resizebox{\linewidth}{!}{
    	\begin{tabular}{|c|c|c|c|c|}
    	\hline
    	Hop  & \multicolumn{2}{c|}{Weekly} & \multicolumn{2}{c|}{Daily}   \\ \cline{2-5}
    	Count & 46 Servers & 8 Servers & 46 Servers & 8 Servers  \\ \hline
    	0 & 47 & 52 & 44 & 58\\ \hline
    	1 & 37 & 46 & 38 & 55\\ \hline
    	2 & 29 & 35 & 29 & 38\\ \hline
    	3 & 15 & 20 & 7 & 24  \\ \hline
    	\end{tabular}
	}
\end{table}

To assess the impact of the length of the reference time period, labels obtained from two subsequent weeks are combined together and the similarity of resulting histogram is evaluated with that of the following week. 
This resulted in a slightly decreased overall similarity as the ratio of rare events has relatively increased. 
To also evaluate changes due to concept drift, similarity scores are measured by skipping weeks between training and testing weeks.
By leaving a gap of eight weeks between the two, we determined that the overall similarity reduced to 30\%.

Another important point is the similarities between servers running the same service.
For 15 groups of servers (out of 31) that include two servers, similarities are evaluated for the same weeks.
This yielded an average similarity of 17\% which shows that despite offering the same service, individual server activity differs noticeably. 
Since file events contribute significantly to the variation in activity in each group, we also examined the similarity only in the context of process activity.
This resulted with an average similarity of 48\% which further supports the idea that these servers operate differently within the enterprise. 

\section{Discussion of Findings}
\label{sec:discussion}

So far we have analyzed the system audit data to understand logon behavior and system execution activity of servers.  
Below, we further examine our findings to guide the development of better analysis methods of server data. 

{\bf {\em Impact of Event Filtering:}}
Filtering of log events during data collection limits our visibility on the overall server activity.
For example, the dataset collected from the real end-user systems in \cite{priotracker} shows that, on average, a machine generates up to millions events per day.
However, this number decreases to around 4 thousand for our dataset, mostly because of the filtering process.
However, the storage requirements for generated logs from potentially thousands of systems force enterprises to accept a trade-off between reducing the utility of log data and maintaining a long-term analysis capability to investigate past incidents.
The problem of reducing system audit data while preserving the forensic utility is in fact an open research problem with some proposed solutions \cite{lee2013loggc,ma2016protracer,xu2016high,hossain2018dependence,hassan2020tactical}.
In practice, however, enterprises rely on public rule sets \cite{SwiftOnSecurity,sigma,sysmonFiletr} or implement their own log collection policies by selectively capturing what they identify as more critical.

In relation to our analysis of log data, the reduction operation cannot induce a rarity in the system. 
That is, if an event is observed once, its later occurrences are guaranteed to be captured. 
Therefore, our findings on rareness of events may only miss some other rare events that were not logged by system utilities.
In the case of similarity of server activity, the elimination of system events prevents us from observing the full scope of provenance relations in the data.
This results in a reduced overall similarity as some common events cannot be taken into account.
Nevertheless, a large degree of dissimilarity among histogram labels obtained from partial provenance graphs shows that the server activity is quite non-stationary.

{\bf {\em Leveraging Logon Statistics:}}
Several studies proposed anomaly detection methods to identify the change in a user's logon activity pattern \cite{nodoze,senator2013detecting,rashid2016new,yuan2020time}.
Since users perform most of their activity on dedicated end-user systems, they perform fewer logons that last longer durations.  
Our findings show that logon patterns on servers vary significantly from this behavior with the frequency of logons changing between once an hour to a week or even a month and each logon session lasting seconds to hours depending on the server type.
The correct modeling of logon behavior on servers is in fact very critical as the shared resources and services hosted by servers are of high interest to attackers. 
As it is evident in the Mitre's enterprise matrix \cite{MITREATT}, several techniques\footnote{Most notably, these include techniques such as T1190, T1606, T1021, T1039, T1114, T1102, T1567, T1505.} relate to attack activity on servers. 
Therefore, it is critical to incorporate the logon behavior of a user with respect to different servers to user-based anomaly detection approaches.
This is also important for synthetic data generators to ensure the creation of realistic background behaviors and to complement efforts such as the CERT Insider Threat Dataset \cite{cert}.

\def\cpar{\hss\egroup\line\bgroup\hss}
\begin{figure*}[!] 
  \begin{minipage}[b]{0.33\textwidth}
    \centering
    \includegraphics[width=1\columnwidth]{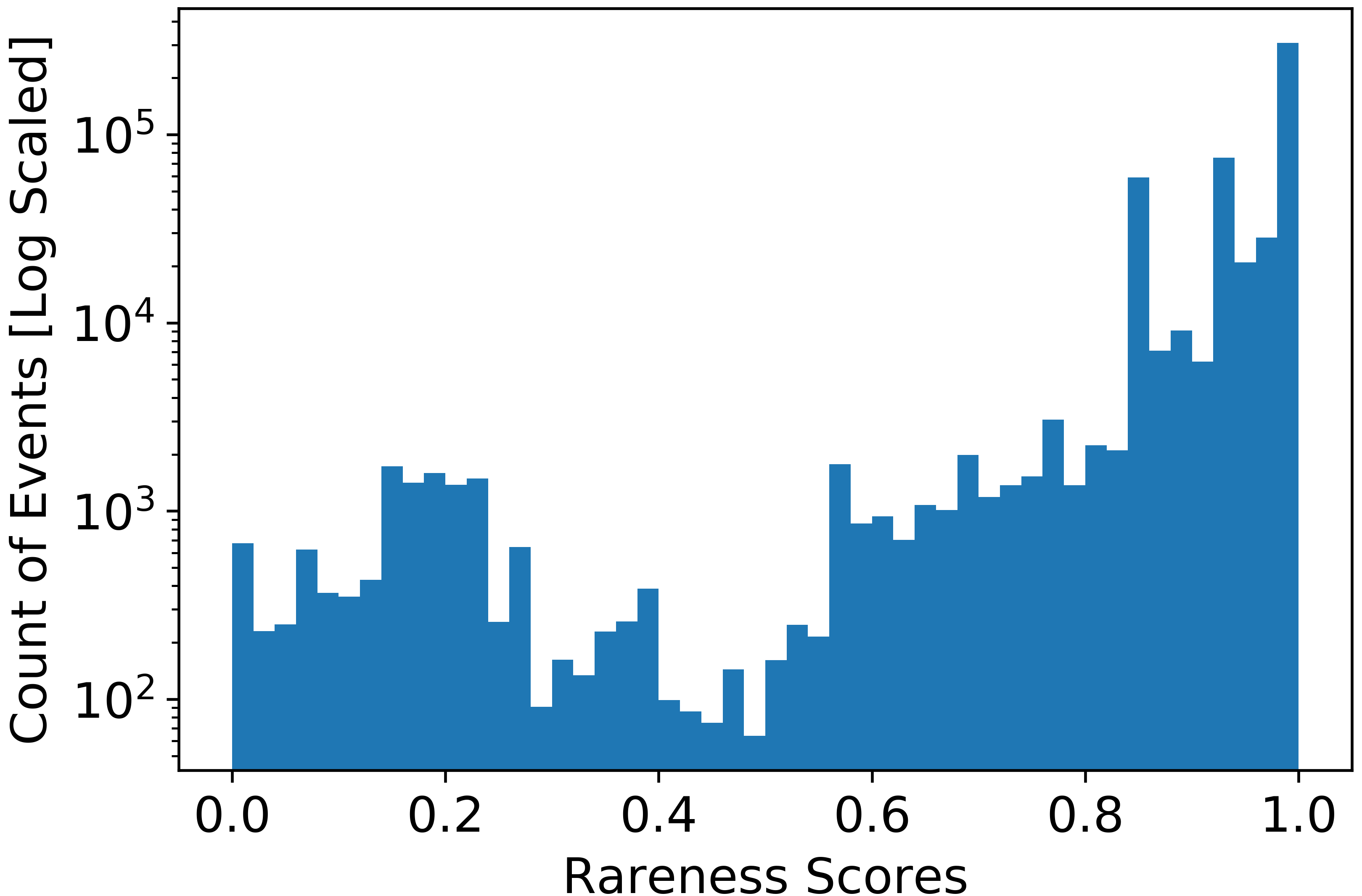}
    \centerline{(a)}
    \label{fig:PrioTrackerAVG}
  \end{minipage}
  \begin{minipage}[b]{0.33\textwidth}
    \centering
    \includegraphics[width=1\columnwidth]{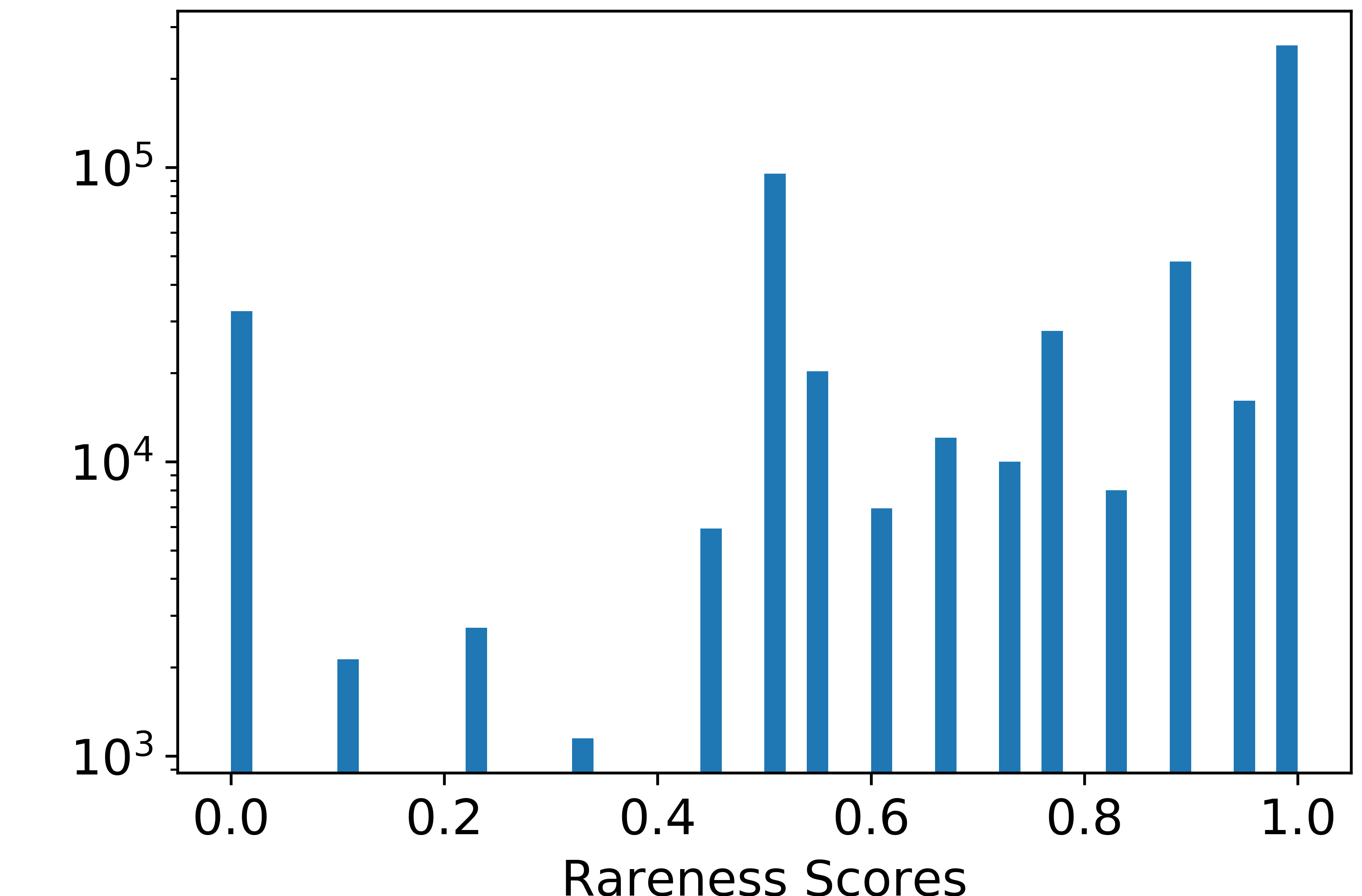}
    \centerline{(b)}
    \label{fig:PrioTrackerDist}
  \end{minipage}
    \begin{minipage}[b]{0.33\textwidth}
    \centering
    \includegraphics[width=1\columnwidth]{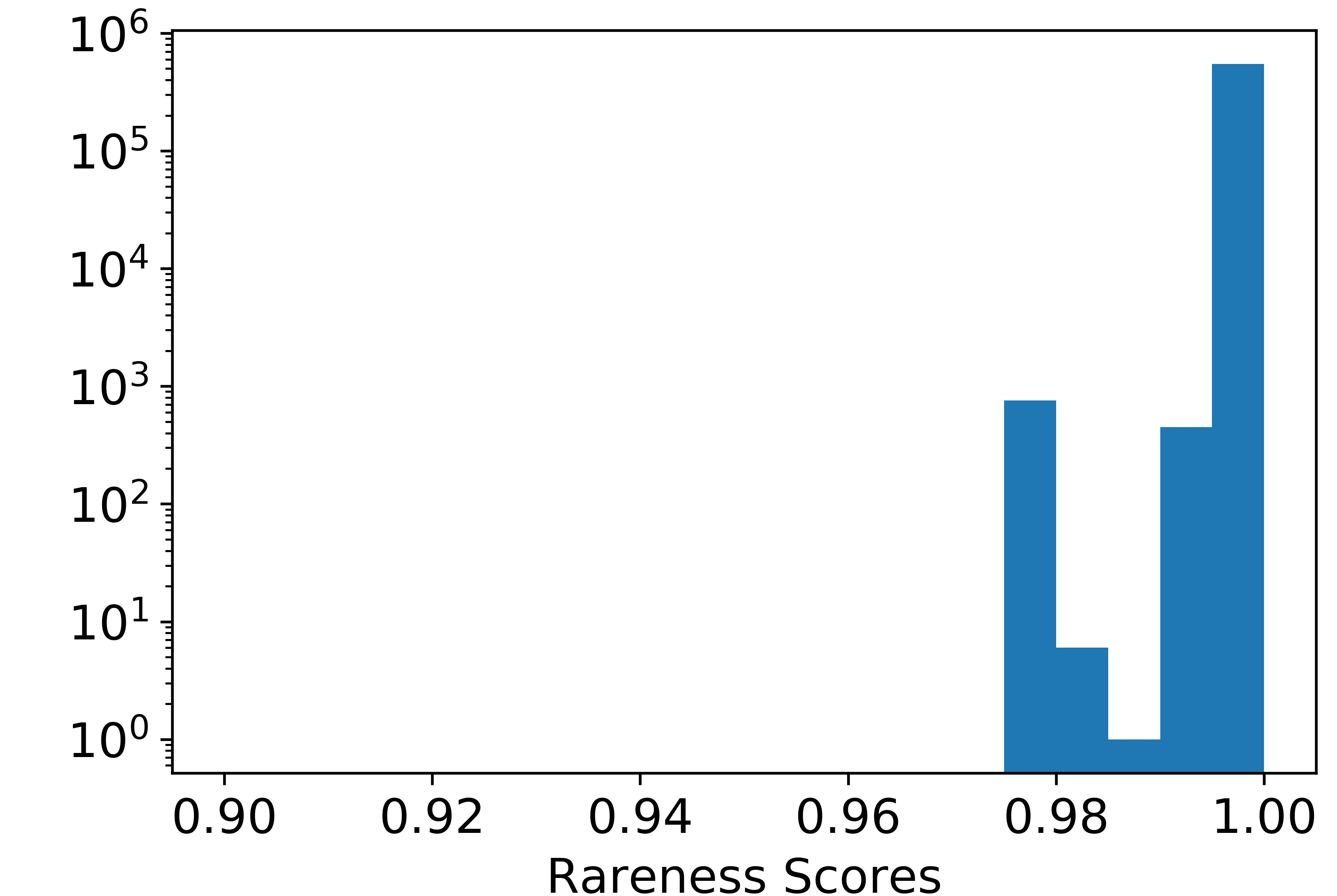}
    \centerline{(c)}
    \label{fig:PrioTrackerDistGroup}
  \end{minipage}
  \vspace{-0.2in}
\caption{Rareness score of events based on different window sizes and grouping approaches. 
(a) Rareness scores are measured daily with grouping similar servers.
(b) Rareness scores are measured weekly with grouping similar servers.
(c) Rareness scores are measured weekly using all servers in the enterprise.}
\label{fig:compare}
\vspace{-0.2in}
\end{figure*}

{\bf {\em Detecting Anomalies in Server Data:}}
Our analysis of the logon behavior of administrator and standard users to servers shows that logons do not exhibit a regularity that can be easily modeled.
This gets more visible on the second category of servers mainly used for innovation and development activity.
It is plausible that user activity on servers is mainly driven by users' daily tasks. 
For example, in our environment, there are (SharePoint) development servers used for developing different applications.
It is observed that some developers access only one of these servers whereas others access any number of them with overall activity changing from week to week. 
We also determined that some development servers are only accessed by administrators in an irregular manner.
Similarly, for some servers especially in the first category, we observed very few logons during the 10-week duration.
This may indicate that administrator users logon to those servers on an as-needed basis. 
These findings crucially reveal that reliable anomaly detection based on user activity must incorporate server activity with other information sources that drive user activity.
In this regard, the incident management systems used by IT teams to monitor and manage servers and the daily tasks of developers can be used to support such decision systems. 

In terms of server activity, the most critical aspect concerns the complexity of audit data for anomaly detection.
That is, does server activity provide a suitable baseline to reliably detect anomalies in system execution?
Our rareness measurements show that a very large number of rare events are encountered on a weekly basis.
This shows that detection of attacks solely based on event-level analyses will yield a huge volume of cases for investigation.
Our results also demonstrate that the similarity of provenance graphs generated daily is around 40\%.
Even when only process-creation events are considered, the similarity increases to 48\%. 
In other words, this implies the presence of a very fast concept drifts in data. 
Overall, it is evident that anomaly detection approaches need to focus on obtaining better contextual representations for events. 
However, due to the diversity of events, this may be a more challenging task than it is assumed to be.

{\bf {\em Duration of Reference Time Period:}}
Rareness scores given in Fig. \ref{fig:PrioTracker}(a) show that when 4-9 weeks are used as a historic time window, the number of rare events decreases slowly with 
the duration of the reference period.
This indicates that further extending the duration of the reference window will potentially reduce the number of rare events.
The ideal duration should be determined as the number of weeks at which a plateau is obtained. 
This issue needs to be further studied on more extensive datasets than ours.

{\bf {\em Resolution of Reference Time Period:}}
Assessing the rareness of events requires splitting the reference time window into intervals, such as days or weeks, when checking the occurrence of a query event as defined in Eqs. \ref{eq:rareness1} and \ref{eq:context}.
In this regard, choosing shorter intervals is expected to yield more rare events. 
For example, if an event occurs on one day of every week, its rareness score will be $0$ and around $0.85$ when the interval is set to a day.
The changes in the distribution of rareness scores computed with respect to Eq. (\ref{eq:rareness1}) are shown in Figs. \ref{fig:compare}(a) and \ref{fig:compare}(b) when the time resolution is set to both days and weeks.
It is evident choosing a longer duration yields a more favorable analysis setting.  

In the case of similarity of the provenance graphs, daily and weekly measurements yielded very similar measurements as presented in Table \ref{MinMAxSim}.
This essentially indicates that a number of events repeat almost on a daily level, thereby creating a baseline of activity that warrants a similar level of activity both at the daily and weekly level.
Events other than those are most likely the rare ones that cause a reduction in the measured similarity.
Hence, applying graph similarity at a daily basis is more preferable as it will make the analysis easier due to the smaller overall graph sizes.

{\bf {\em Server/Group Level or Enterprise-wide Assessment:}}
Another related issue concerning computation of rareness scores is to consider a single server or all available servers when evaluating the occurrence frequency of events.
Extending the event context to all servers will in fact contribute to the rareness of events, especially for those that are specific to a server.
This is because such events will be designated frequent when evaluated locally but will appear as rare when considered in the global context as the ratio of the servers that encountered those events will be low. 
At the same time, however, evaluation of the frequency of events using data from multiple systems with a similar setup will help more accurately identify rare events.
This can be realized by grouping together servers that are expected to operate similarly.
Figures \ref{fig:compare}(b) and \ref{fig:compare}(c) demonstrate the change in distribution of rareness scores when only servers in the same service group and all the servers are taken into account.
Further, the fact that all events are assigned a high rareness score in Fig. \ref{fig:compare}(c) also shows that  
there are really few common events between different types of servers.

To identify the appropriate grouping scheme for servers, we computed average rareness scores for different event types from servers in an individual manner,
by grouping them based on the service they provide and their location, and by allowing the group to include all servers in the enterprise.
As can be seen in Table \ref{tab:grouping}, increasing the number of servers in the group results in an increase in the average rareness scores of events encountered on a server.
This finding is consistent with our previous result that common events are rare between different servers.
Hence, when the grouping is completely removed and the data from each server is analyzed by isolating the reference time window of each server to historic data from itself, the lowest rareness scores are obtained.
However, this basic setting makes poisoning attacks easier because attackers can repetitively perform certain tasks on a server they control to hide their activities.
Therefore, it is more desirable to group together servers that run the same service under similar conditions together, striking a balance between the ability detect poisoning attacks and the number of rate events.

\begin{table}[!ht]
	\centering
	\caption{Average rareness scores for different event types obtained by applying different grouping approaches}
	\label{tab:grouping}
	\resizebox{\linewidth}{!}{
    	\begin{tabular}{|c|c|c|c|c|}
    	\hline
    	Grouping Approach & \multicolumn{4}{|c|}{ Avg. of Rareness Scores }   \\ \cline{2-5}
    	 (No. groups)  & File & Process & Network & Registry  \\ \hline
    	Server Level (46)& 0.66 & 0.37 & 0.66 & 0.91\\ \hline
    	Same Type \& Location (31) & 0.78 & 0.57 & 0.76 & 0.93\\ \hline
    	Same Type (23)& 0.88 & 0.68 & 0.82 & 0.95\\ \hline
    	All Servers & 0.99 & 0.98 & 0.99 & 0.99  \\ \hline
    	\end{tabular}
	}
\end{table}

\section{Conclusions}
In this work, we perform the first feasibility study of utilizing server activity in an enterprise to detect anomalies. From a corpus of 10-week enterprise network logs, we analyze the server logon activities, the rareness of observed server events, and the similarity of the provenance graphs for system activities. Our observations show that servers, in general, exhibit high variation in their behavior. However, careful consideration of training parameters and grouping of servers, we show that the rareness measurement improves by 24.5\% on average with our data. Further, we observe an improvement of 5.5\% on average in the similarity of provenance graphs when profiled daily compared to weekly basis. While some of the results are encouraging, we believe more needs to be done in order to improve the rareness score and provenance similarity, potentially utilizing unfiltered data as well as additional contextual information, in order to complement existing user based enterprise anomaly detection techniques.

\bibliographystyle{IEEEtran}
{\footnotesize \bibliography{bibfile}}

\end{document}